\documentclass[aps,apl,showpacs,twocolumn,amsmath,amssymb]{revtex4}

\usepackage{graphicx}
\usepackage{dcolumn}
\usepackage{bm}

\begin{document}

\title{Relaxation and Emittance Growth of a Thermal Charged-Particle Beam}

\author{Tarc\'isio N. Teles}

\email{teles@if.ufrgs.br}

\author{Renato Pakter}

\email{pakter@if.ufrgs.br}

\author{Yan Levin}

\email{levin@if.ufrgs.br}

\affiliation{Instituto de F\'isica, UFRGS, Caixa Postal 15051, CEP 91501-970 Porto Alegre, Rio Grande do Sul, Brazil}

\begin{abstract}
  We present a theory which allows us to accurately calculate the distribution functions and the emittance growth
  of a thermal charged-particle beam after it relaxes to equilibrium. 
  The theory can be used to obtain the fraction of particles which will evaporate from the beam to form a halo.  
  The calculated emittance growth is found to be in excellent agreement with the simulations.
\end{abstract}

\maketitle

The understanding of physics involved in the transport of high-intensity charged-particle beams 
is of fundamental importance in the development of a new generation of accelerators and electromagnetic 
wave generators to be used in applications such as heavy ion fusion, high-energy physics, communication, 
materials processing, and cancer therapy. A very detrimental effect that may seriously influence the 
efficiency of such devices is a halo formation and emittance growth of the beam 
\cite{ref:gluckstern,banna02,wangler02,brown05,mugli08}. These not only 
cause degradation of the beam quality, but may also be responsible for the activation 
of accelerator channel wall and pulse shortening in microwave devices. Emittance growth is 
generally associated with the relaxation of initially non-stationary beam towards a more stable stationary 
configuration.  The emittance growth can be calculated if the final stationary distribution is 
known \cite{reiser,wangler02,ref:roger}. However, the determination of this distribution 
is not an easy task~\cite{reiser,ref:davidson,zhou} because particles in an intense beam interact through long-range forces which 
prevent the system from relaxing to the true 
thermodynamic equilibrium~\cite{ref:padmanaban,ref:chavani2006,ref:ruffoprl08}.   
Instead these systems
get trapped in metastable states,  lifetime of which diverges with the number of particles.   
To understand the properties of these states, one can not use the standard statistical mechanics and
new non-equilibrium theories must be developed \cite{ref:levinprl}. 

In this letter, we will present a theoretical framework which will allow us to 
accurately calculate the density and the
velocity distributions of particles in the 
final stationary state achieved by a space-charge dominated beam focused by a uniform external magnetic field. 
Our approach is based on the theory of violent relaxation in gravitational systems~\cite{Ly67}, modified so as to
explicitly account for 
the effects of single particle resonances \cite{ref:gluckstern}, 
responsible for the halo formation \cite{ref:levinprl}.
The theory is applicable to arbitrary initial conditions.  In this letter we will show how the theory
can be used to accurately calculate the density and the velocity distributions, as well as, to  account 
for the emittance growth of a charged-particle beam 
launched with a thermal (Maxwell) velocity distribution.
The predictions of the theory will be tested against the molecular dynamics simulations.

The physical system considered here is an intense charged-particle 
beam of perveance $K=2q^2N_b/\gamma_b^3 v_z^2 m$ --- 
where $c$ is the speed of light {\it in vacuo}, and $q$, $m$, and $\gamma_b=[1-(v_z/c)^2]^{-1/2}$ are the charge, 
mass, and the relativistic factor of the beam particles, respectively --- propagating with an axial velocity $v_z\hat {\bold e}_z$ 
through a magnetic focusing
channel enclosed by a cylindrical conducting wall located at $r=r_w$ \cite{ref:reiser,ref:davidson}. 
The external focusing magnetic field is given by $\bold B=B_0\hat{\bold e}_z$. 
It is convenient to work in the Larmor frame, 
which rotates with respect to the laboratory frame with angular velocity $\Omega_L=qB_0 /2\gamma_b v_z m c$, normalized to $v_z$. 
In the Larmor frame, the external magnetic field produces
a parabolic confining potential $U(r)=\kappa_z r^2/2$, 
with the focusing field parameter $\kappa_z=\Omega_L^2/c^2$. 
The effective electromagnetic scalar potential between the particles  $\psi$, 
incorporates both the self-electric and the self-magnetic fields, $\bold
E^s$ and $\bold B^s$. This potential satisfies the 
Poisson equation with the boundary condition  $\psi(r_w)=0$, 
\begin{equation}
  \nabla^2\psi=-(2\pi K/N_b)n_b(\bold r,s) \label{eqpoisson}
\end{equation}
where, $N_b$ is the number of particles per unit axial length, $\bold r$ is the position vector in the transverse plane, and
$n_b(\bold r,s)=N_b\int f d^2\bold v$ is the transverse beam density profile, given in terms of the one particle distribution
function $f(\bold r,\bold v;s)$. In the  Larmor frame, the dynamics of the beam reduces to that of 
a two dimensional one component plasma with logarithmic interaction between the particles, 
confined by a parabolic potential $U(r)$.  
The axial coordinate $s = z = v_zt$ plays the role of time for this two dimensional system.   

We will suppose that the initial (transverse) distribution of the beam is Gaussian in velocity space and is uniform in crossection,
\begin{equation}
  f_{0}(\bold r,\bold v)=\frac{1}{2\pi^2\sigma^2r_m^2}\Theta(r_m-r)e^{-(v^2/2\sigma^2)},
\label{gauss}
\end{equation}
where $\sigma^2$ is the initial mean square transverse velocity, and $r_m$ is the beam radius.  
The quality of the beam 
is inversely proportional to
the emittance, defined as $\varepsilon^2 = 4<r^2><v^2>$, for a {\it stationary} beam.  
For the distribution (\ref{gauss}), the emittance is $\varepsilon_0=2 \sigma r_m$.

It will be convenient to discretize Eq. (\ref{gauss}) into a {\it p}-level distribution
\begin{equation}
f_{p}^0(\bold r,\bold v)=\sum_{j=1}^p\eta_j \rho^0_j(\bold r,\bold v)\;, \label{gausdiscre}
\end{equation}
where $\rho^0_j(\bold r,\bold v) \equiv \Theta(v-v_{{j-1}})\Theta(v_{j}-v)\Theta(r_m-r)$, and 
$v_{j}$ and $\eta_j$ are the maximum velocity and the amplitude of the level $j$, respectively, with  $v_0=0$. 
For a perfect description of Eq. (\ref{gauss}), an infinite number of levels ($p\to\infty$) in (\ref{gausdiscre}) will be necessary.  
In practice, however, we find that a small number of levels is already sufficient to provide a very accurate approximation for the beam dynamics. 
For a given value of {\it p}, the optimal values of $\eta_j$ and $v_{j}$ can be obtained 
by minimizing the functional ${\cal F}=\int(f_{0}-f_{p}^0)^2d^2\bold rd^2\bold v$, with the  constraints on the  kinetic 
energy and normalization,
\begin{eqnarray}
\delta\{{\cal F} +\lambda_1(\int \frac{v^2}{2}f_{p}^0d^2\bold r d^2\bold v
- \sigma^2) + \nonumber \\ + \lambda_2( \int f_{p}^0d^2\bold r d^2
\bold v -1)\}=0\;,
\label{disc}
\end{eqnarray}
where $\lambda_1$ and $\lambda_2$ are the two Lagrange multipliers.  Minimization of Eq.~\ref{disc} yields 
the optimal parameters $\{\eta_j\}$ and $\{v_j\}$.  The many-body dynamics of systems with unscreened 
long range interaction is governed by the collisionless Boltzmann (Vlasov) equation. The 
distribution functions which satisfy the Vlasov equation evolve in time as the density of an incompressible fluid. In particular
this means that the $p$ hyper-volumes, $\gamma(\eta_j)=\int \delta(f_p^0({\bf r},{\bf v})-\eta_j) {\rm d^d{\bf r}}{\rm d^d{\bf v}}$,  
of the distribution (\ref{gausdiscre}) will be preserved by the Vlasov flow \cite{Ly67}. 
 
In reference \cite{Ly67} it was argued that the 
stationary solution of the Vlasov equation could be obtained by maximizing the coarse grained entropy, with the constraints imposed
by the conservation of energy and the hyper-volumes of the $p$ levels of the initial distribution function. For matched beams and 
water-bag initial conditions,
the resulting distribution was shown to be in excellent agreement with the 
molecular dynamics simulations  \cite{ref:levinprl}. However, for mismatched beams, the plasma oscillations result in parametric
resonances, which lead to a significant particle evaporation.  After the relaxation process is complete, the stationary
beam phase separates into a cold core, surrounded by a halo of highly energetic particles. 
For a water-bag initial condition ($p=1$),
it was shown that the core was very well described by a cold Fermi-Dirac distribution with the temperature $T \approx T_F/40$, 
where $T_F$ is the ``Fermi temperature'' of the beam. 
The halo was reasonably approximated by a step function, with energy range of one particle resonance. 
The full distribution function had the form of
\begin{equation}
f(\bold {r, v})= f_{c}(\bold {r, v}) + f_{h}(\bold {r, v})\;. 
\label{f}
\end{equation}

For a $p$-level system, which is used to  approximate the  thermal distribution 
given by Eq. (\ref{gauss}),  a similar phase separation will occur.  
The form of the core distribution function can be obtained, once again, by
maximizing the coarse grained entropy to yield 
\begin{equation}
f_{c}(\bold {r, v})=\sum_{j=1}^p (\eta_j-\chi) \rho_{j}(\bold r, \bold v) \;,
\label{fc0}
\end{equation}
with
\begin{equation}
\rho_j(\bold {r, v})=
\frac{e^{-\beta \eta_j\epsilon(\bold{r,
v})+\alpha_j}}{\sum_{i=1}^pe^{-\beta \eta_i \epsilon(\bold r,\bold v) +
\alpha_i}+1}\;,
\label{fc}
\end{equation}
where the mean particle energy is $\epsilon(\bold r,\bold v)=v^2/2 +U(r)+\psi(r)$, and  $\beta$ and  
$\{\alpha_j\}$  are the Lagrange multipliers for the energy and the hyper-volumes conservation. The oscillations of the 
mismatched beam excite the  parametric resonances resulting in a halo formation \cite{ref:levinprl}. 
The parameter $\chi$ determines
the fraction of the particles which will evaporate to form the halo of the beam. 
The coarse-grained distribution can no longer
preserve all the hyper-volumes of the original fine-grained distribution function, so that only the 
lower energy hyper-volumes will be conserved,
while the particles from the higher energy states will evaporate to form a halo.   
We find that the halo can be modeled accurately
by the distribution    
\begin{eqnarray}
&&f_{h}(\bold {r, v})=\chi\Theta(\epsilon_{\zeta}-\epsilon(\bold {r, v}))+ \nonumber \\
&&\chi\Theta(\epsilon(\bold {r, v})-\epsilon_{\zeta})\Theta(\epsilon_{R}-\epsilon(\bold {r, v}))e^{-\gamma(\epsilon -\epsilon_{\zeta})}\;
\label{fh}
\end{eqnarray}
The extent of the halo \cite{ref:gluckstern} is  up to one particle 
resonance energy, $\epsilon_{R}$.  The low energy part of the halo distribution is flat,
while for energies $\epsilon > \epsilon_{\zeta} = \epsilon_{R}/2$, it decays  exponentially with exponent  
$\gamma \approx 8$.  We can now, in principle, numerically solve Eqs.(\ref{eqpoisson}, \ref{f}, \ref{fc}) and (\ref{fh}) to calculate the the
stationary distribution function $f(\bold {r, v})$ of the relaxed beam.  
There is, however, one problem.  Equations (\ref{fc}) and (\ref{fh}) contain $p+2$ parameter:
$\beta, \{\alpha_j\}$, and $\chi$.  The conservation of energy, norm and of lower energy hyper-volumes gives us $p+1$ additional equations.
\begin{eqnarray}
\int d^2\bold r d^2\bold v \epsilon(\bold{r, v})f(\bold{r, v})=\epsilon_0 \;,\nonumber \\ 
\int d^2\bold r d^2\bold v f(\bold{r, v})=1 \;, \\
\int d^2\bold r d^2\bold v \rho_{j}(\bold{r, v})=\int d^2\bold r d^2\bold v \rho_j^0(\bold{r, v}) \;,\nonumber  \label{constrains}
\end{eqnarray}
where $1 \le j \le p-1$  and  $\epsilon_0$ is the average energy per particle of the initial thermal distribution,
\begin{equation}
\epsilon_0=\sigma^2+\kappa_z\frac{r_m^2}{4}+K\left[\frac{1}{8}-\frac{1}{2}\ln\left(\frac{r_m}{r_w}\right)\right] \;.
\end{equation}
There, however, still remains one missing condition necessary to uniquely determine the distribution function.  For
water-bag distributions, this condition was provided by the requirement that in the relaxed state, the core temperature is
very low, $T \approx T_F/40$.   It is difficult, however, to numerically implement this condition for $p$-level distributions. 
On the other hand, if we discretize the original thermal distribution into only one level ($p=1$),
the condition $T \approx T_F/40$ is easily implemented and 
allows us to uniquely close all the equations and calculate the relaxed distribution function  \cite{ref:levinprl}.
We find, that although the core distribution is not well described  by a $p=1$ system, the halo part of the distribution  
is found to be quite accurate.  This allows us 
to fix the value of $\chi$.   Using this  $\chi$, we can now improve the
description of the core region by including additional levels into the discretization procedure.  
To compare the predictions of the theory with the simulations, we calculated the 
number of particles in the interval  
$[r,r+dr]$, $N(r)dr=2\pi N_b
rdr\int d^2\bold vf(\bold{r, v})$, and the number of particles with velocities between $v$ and $v+dv$, $N(v)dv=2\pi N_bvdv\int
d^2\bold rf(\bold{r, v})$, for various initial conditions.  
The simulations are based on the Vlasov dynamics in which
particles interact with the mean-field potential.  This avoids the 
collisional effects  present in finite size systems, but which must
vanish for  one component plasmas in thermodynamic limit.  
The simulation code uses the  Gauss law to calculate the mean
electric field felt by each particle~\cite{ref:roger}. For axisymmetric
beams studied in this work, this proves to be very efficient 
since the electric field at a radial coordinate $r$ is
determined simply by counting the total number of particles 
with coordinates smaller than $r$. Simulations were performed with $20000$ particles.
As can be seen from the Figs. 1 and 2, the agreement between the
theory and the simulations is excellent.  In the figures, the distances are measured in units of $\sqrt{\varepsilon_0 /\Omega_L}$
and the velocities are in units of $\sqrt{\varepsilon_0 \Omega_L}$. We have also defined a scaled perveance $K^* \equiv K/\Omega_L \varepsilon_0$ and 
the mismatch parameter,  $\mu \equiv r_m/r_0$, which measures the deviation of the initial beam radius 
from the corresponding virial value $r_0= \sqrt{K+4 \sigma^2}/\Omega_L$, for which the oscillations of the beam envelope are very small. 
In particular, we find that the discretization of the Gaussian by only 4-levels, 
already provides us with an almost perfect description of the core region.

\begin{figure}[!htb]
\begin{minipage}{0.45\textwidth}
\begin{center}
\includegraphics[width=5.cm,height=5.cm,angle=0]{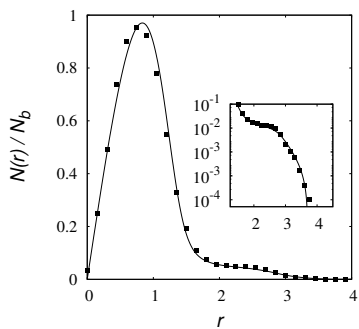}
\end{center}
\end{minipage}
\caption{The relaxed particle density of an initially thermal beam with scaled perveance $K^*=1$ and mismatch of $75 \%$ ($\mu=1.75$). 
The points are the results of the simulations and the solid line is the prediction of the theory. 
Inset shows the exponential decay of the halo close to one particle resonance energy.}
\label{fig1}
\end{figure}

\begin{figure}[!htb]
\begin{minipage}{0.45\textwidth}
\begin{center}
\includegraphics[width=6.cm,height=3.cm,angle=0]{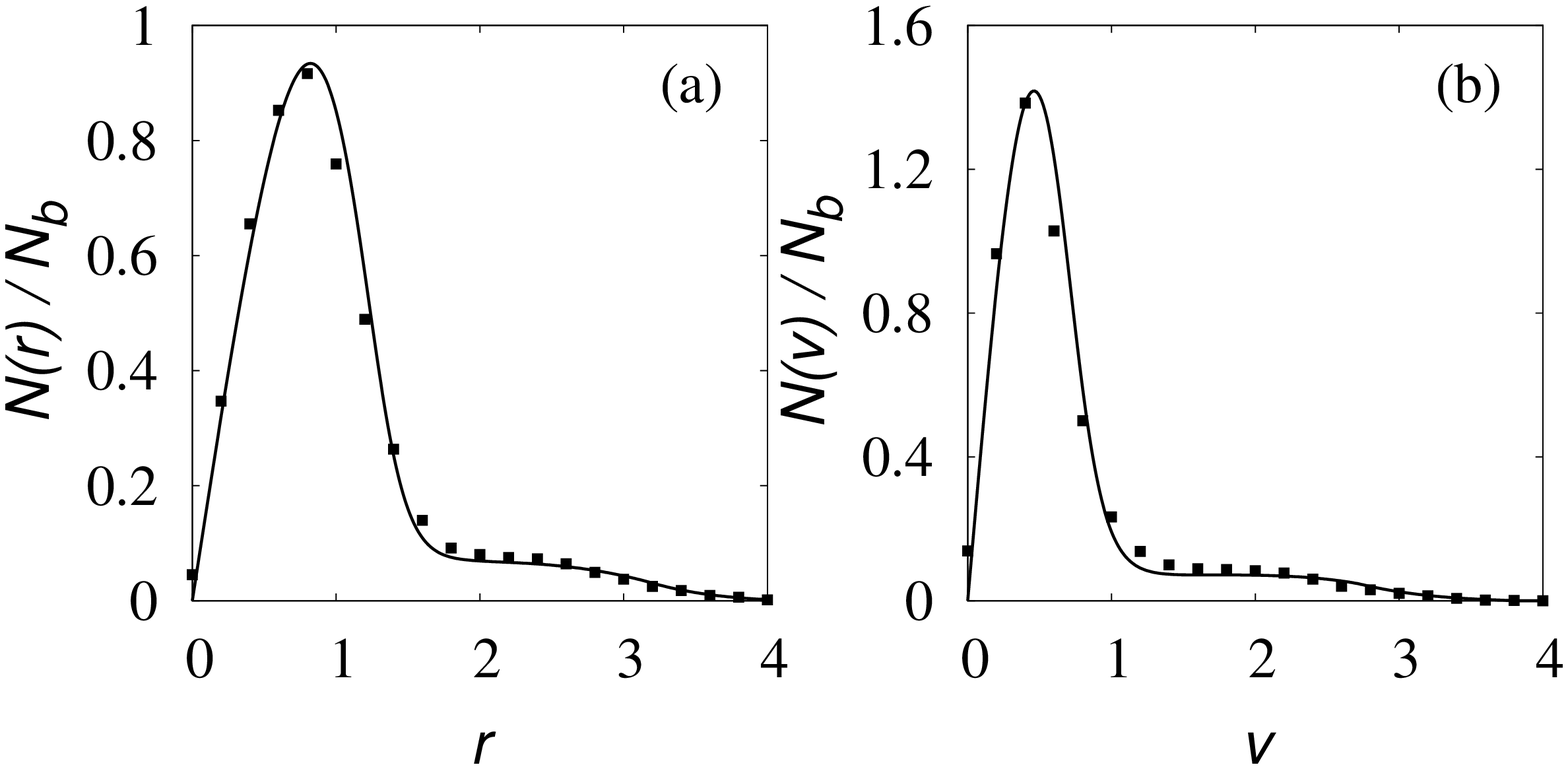}
\end{center}
\end{minipage}
\caption{The density and velocity distributions: solid line is the prediction of the theory and points are the 
results of the molecular dynamics simulation. The scaled perveance is $K^*=1$ and mismatch is $50 \%$ ($\mu=1.50$).}
\label{fig2}
\end{figure}

As a direct application of the theory developed above, we calculate the emittance  growth of an originally thermal beam.  This
quantity is of fundamental importance for the design and development of high intensity space-charge dominated 
beams {\cite{ref:wang98}.
The calculations are performed for beams of varying scaled perveance $K^*$ and mismatch parameter $\mu$.
The results are compared with the molecular dynamics simulations.  Once again, an excellent agreement is found between the theory
and the simulations (Fig. 3).

\begin{figure}[!ht]
\begin{minipage}{0.45\textwidth}
\begin{center}
\includegraphics[width=5.cm,height=5.cm,angle=0]{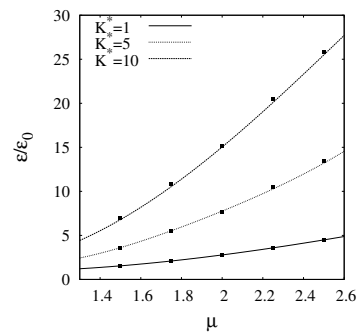}
\end{center}
\end{minipage}
\caption{The final emittance of an initially thermal beam. The points are the result of molecular dynamics simulation 
and the lines are the predictions of the theory.}
\label{fig3}
\end{figure}

To conclude, we have presented a theory which allows us to calculate the  density and the velocity distributions of an initially thermal
beam after it relaxes to the final stationary state.  Comparing to the simulations, the theory is found to be 
extremely accurate, without any adjustable parameters.  In particular, it can be used to calculate the emittance growth and the
fraction of  particles which will evaporate  as the beam evolves to its final stationary state.

This work is supported by CNPq, FAPERGS, and INCT-FCx of Brazil, and by the Air Force Office of Scientific Research 
(AFOSR), USA, under the grant FA9550-09-1-0283.

\clearpage
\appendix
\end{document}